\documentclass[aps,pra,reprint,superscriptaddress,nofootinbib,longbibliography]{revtex4-2}

\usepackage{amsmath,amssymb,bm,mathtools}
\usepackage{graphicx}
\usepackage{booktabs}
\usepackage{microtype}
\usepackage{hyperref}
\usepackage{xcolor}
\usepackage{siunitx}
\usepackage{amsthm}
\usepackage{orcidlink}

\hypersetup{colorlinks=true,linkcolor=black,citecolor=black,urlcolor=blue}
\graphicspath{{figures/}}

\newcommand{\Tr}{\operatorname{Tr}}
\newcommand{\deff}{d_{\rm eff}}

\begin{document}

\title{Readout Orientation Controls Measurement-Accessible Quantum Tangent Geometry}

\author{Marwan Ait Haddou\,\orcidlink{0009-0008-1734-1721}}
\email{aithaddou.marwan@outlook.com}
\affiliation{Independent Researcher, Morocco}

\begin{abstract}
A fixed quantum measurement can expose substantially more tangent information than a restricted observable readout retains. We study this second restriction directly. For a normalized covariance $C\succeq0$, $\Tr C=1$, of measurement-induced tangent scores in an $N$-dimensional centered score space, and a rank-$r$ readout projector $P$, we quantify retained tangent mass by $R=\Tr(PC)$. The ratio $\rho=R/(r/N)$ separates the actual retained mass from a rank-only random-orientation reference. Standard Grassmann averaging gives $\mathbb{E}\rho=1$ and
\begin{equation*}
\mathrm{Var}(\rho)=\frac{2(N-r)[N\Tr(C^2)-1]}{r(N-1)(N+2)}
\le \frac{2}{r\deff},
\qquad \deff=\frac{1}{\Tr(C^2)}.
\end{equation*}
We use this identity as a null model rather than as a new random-projection theorem. Numerically, family-balanced one- and two-body readouts remain close to the rank reference through $n=16$ even as the tangent covariance becomes strongly anisotropic. The decisive equal-rank comparison then holds the circuit, measurement record, readout rank, and evaluation shot budget fixed. For Haar-$U(4)$ at $n=12$, cross-fitted alignment increases the mean directional gradient-energy proxy by a factor $9.584$ and the finite-shot signal-to-noise ratio by a factor $3.111$ relative to the physical one-body readout, while a random rank-matched subspace remains near the rank baseline. A half-filled $U(1)$-conserving family provides a structured counterexample to generic orientation: physical low-weight $Z$ readouts are already strongly aligned with leading tangent directions over the tested finite-size range. We treat this symmetry result as a case study, not as a claim that $U(1)$ symmetry generically prevents barren plateaus or that hydrodynamics is the established mechanism. The results isolate readout orientation as a degree of freedom that is invisible to rank alone but directly controls how much measured tangent information remains usable after readout restriction.
\end{abstract}

\maketitle

\section{Introduction}

Variational quantum circuits are ultimately accessed through measurement outcomes. State-space quantities such as the quantum Fisher information (QFI), quantum geometric tensor, and natural-gradient metric characterize infinitesimal distinguishability before a particular measurement and readout are fixed \cite{BraunsteinCaves1994,Stokes2020,Haug2021}. Once a measurement is chosen, the QFI contracts to a classical Fisher geometry; once only a restricted family of observables or features is retained, the experimentally available geometry contracts again \cite{BraunsteinCaves1994,HottaOzawa2004,LuLuoOh2012}. The second contraction is the focus of this work.

This question is adjacent to, but distinct from, several established trainability results. Observable locality can change gradient concentration in parametrized quantum circuits, circuit controllability and dynamical Lie-algebra structure can alter barren-plateau behavior, and symmetry-restricted ansatzes can explore much smaller invariant sectors than generic expressive circuits \cite{Cerezo2021,UvarovBiamonte2021,Larocca2022,HolmesSharma2022}. Random measurements can also approximate quantum Fisher geometry after averaging over measurement bases \cite{LuSha2025}. These results establish that measurement choice, observable structure, and circuit architecture matter. Our question is narrower: \emph{after the quantum measurement has been fixed, and after the dimension of the retained readout has been fixed, how much tangent information is accessible to that particular readout subspace?}

The distinction matters because readout dimension does not determine readout usefulness. Consider the centered score space generated by a fixed measurement. Let $C\succeq0$, $\Tr C=1$, denote the covariance of normalized tangent-score vectors, and let $P$ be the orthogonal projector onto a retained observable subspace of rank $r$. We define
\begin{equation}
R(P,C)=\Tr(PC),
\qquad
\rho(P,C)=\frac{\Tr(PC)}{r/N}.
\label{eq:def-rho}
\end{equation}
Here $R$ is the fraction of normalized tangent mass visible to the readout. The denominator $r/N$ is the mean overlap of a rank-$r$ subspace with a trace-one covariance under random relative orientation. It is therefore a rank-only reference, not a model of a physical circuit family.

The random-subspace identity itself is standard Grassmann geometry \cite{CollinsMatsumoto2017,Bendokat2024,Fernandez2025}. Our earlier manuscript derived the corresponding exact readout-rank law under joint state--tangent isotropy, including the finite-size Beta distributions for the successive measurement and readout projections \cite{AitHaddou2026IsotropicRankLaws}. The present work starts from the failure mode identified there and removes isotropy as an assumption. We use it for a different purpose: to separate three ingredients that are otherwise easily conflated. The rank $r$ fixes the null accessibility scale $r/N$; the spectrum of $C$ controls the width of random-orientation fluctuations; and the relative orientation of $P$ and the eigenspaces of $C$ determines the actual retained mass. In particular, concentration of $\rho$ around one does not imply that $C$ is isotropic. If
\begin{equation}
\deff(C)=\frac{1}{\Tr(C^2)},
\label{eq:deff-intro}
\end{equation}
then the variance of $\rho$ is bounded by $2/(r\deff)$. Thus rank-typical overlap only requires the product $r\deff$ to be large; the ratio $\deff/N$ may still be small.

We test this separation in several circuit families using computational-basis measurement and low-weight diagonal $Z$ readouts. The numerical results have four parts. First, family-balanced generic circuits are close to the rank-only reference for one- and two-body readouts over the tested sizes, although their tangent covariance becomes strongly anisotropic. Second, architecture-resolved deviations occur on both sides of the reference, showing that the generic result is not an isotropy statement and not a claim that physical circuits become Haar-oriented. Third, a rank-matched spectral comparison shows that generic circuits contain leading tangent subspaces that retain much more mass than the physical low-weight span. This establishes that low physical retention can be an orientation mismatch rather than an absence of tangent information at that rank. Fourth, the same-rank distinction survives an operational test: at fixed circuit, measurement record, rank, and shot budget, cross-fitted aligned readouts yield larger directional gradient energy and larger finite-shot signal-to-noise ratio than the physical readout, while random rank-matched subspaces behave near the rank baseline.

A half-filled $U(1)$-conserving RZ--XY circuit family is included as a structured case study. Its low-weight diagonal readout is already strongly aligned with the leading tangent subspace, so the gain from replacing it with a learned same-rank spectral subspace is much smaller than in generic Haar-$U(4)$ circuits. This observation is consistent with several known mechanisms---symmetry-restricted controllability, fixed-charge geometry, readout compatibility with local charge, and slow operator structure under conservation laws---but the present paper does not identify which mechanism dominates. We therefore do not claim that $U(1)$ symmetry generically improves trainability, nor do we infer a hydrodynamic exponent from the finite-size data.

The resulting message is deliberately limited:
\begin{center}
\emph{Rank fixes a baseline; spectral orientation controls actual retained tangent information.}
\end{center}
This is a statement about measurement-accessible tangent geometry. It is not, by itself, a barren-plateau theorem for a supervised loss. Task-level trainability additionally depends on the encoding, labels, classical postprocessing, loss function, and optimization trajectory.

\begin{figure*}[t]
\includegraphics[width=\textwidth]{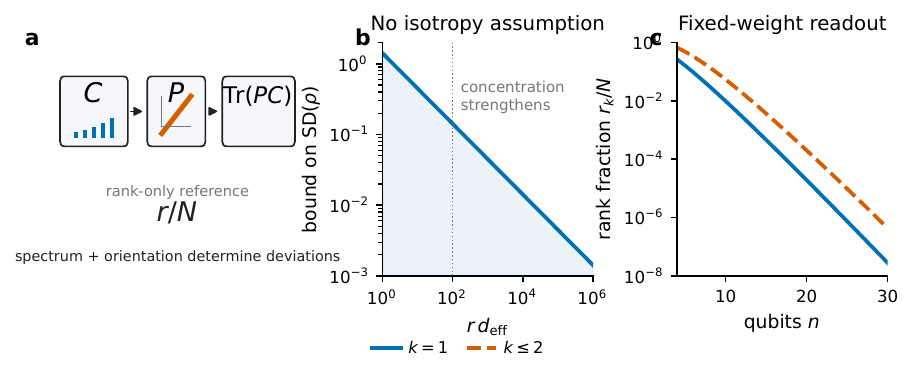}
\caption{Rank and orientation in measurement-accessible tangent geometry. A fixed quantum measurement induces a centered score space. A normalized tangent covariance $C$ distributes tangent mass within that space, while a rank-$r$ readout projector $P$ selects the directions retained by the learner. The accessible tangent mass is $R=\Tr(PC)$. Random relative orientation gives the rank-only reference $r/N$; the spectrum of $C$ controls the null width, while the physical orientation of $P$ controls the observed deviation from that reference.}
\label{fig:framework}
\end{figure*}

\section{Measurement-accessible tangent geometry}

\subsection{From a fixed measurement to normalized tangent scores}

Let $p_\theta(x)$ be the outcome distribution of a fixed measurement applied to a parametrized quantum state $\rho_\theta$. For a tangent direction $v$ in parameter space, the directional derivative $\partial_v p_\theta(x)$ defines a tangent to the classical probability simplex. After removing the constant direction and applying the Fisher normalization used in the numerical protocol, each regular tangent is represented by a centered score vector $u_v$ in an $N$-dimensional real score space. We normalize regular directions so that
\begin{equation}
\|u_v\|_2^2=1.
\end{equation}
Averaging over the tangent ensemble gives
\begin{equation}
C=\mathbb{E}_v\left[u_vu_v^{\mathsf T}\right],
\qquad
C\succeq0,
\qquad
\Tr C=1.
\label{eq:C-def}
\end{equation}
The eigenvectors of $C$ identify score-space directions that repeatedly carry tangent variation, while its eigenvalues quantify how strongly that variation is concentrated.

The effective spectral dimension
\begin{equation}
\deff(C)=\frac{1}{\Tr(C^2)}
\label{eq:deff}
\end{equation}
is maximal at $N$ for the isotropic covariance $I/N$ and decreases as tangent mass concentrates into fewer directions. We also use the normalized purity
\begin{equation}
N\Tr(C^2)=\frac{N}{\deff},
\label{eq:purity}
\end{equation}
which equals one only in the isotropic case.

\subsection{Restricted readout as a score-space projector}

A retained family of observables or linear score features spans a subspace of the centered score space. Let $P=P^{\mathsf T}=P^2$ project onto that span and let
\begin{equation}
r=\Tr P.
\end{equation}
For a normalized tangent $u_v$, the retained squared score norm is $\|Pu_v\|_2^2$. Averaging over tangent directions gives
\begin{equation}
\mathbb{E}_v\|Pu_v\|_2^2
=\Tr\!\left(P\,\mathbb{E}_v[u_vu_v^{\mathsf T}]\right)
=\Tr(PC)
\equiv R(P,C).
\label{eq:R-operational}
\end{equation}
Thus $R$ is not introduced as an abstract overlap alone: it is exactly the mean fraction of normalized score energy retained by the readout.

This formulation separates the measurement from the readout restriction. The full measurement record determines the score geometry from which $C$ is constructed. The projector $P$ then describes which linear combinations of that already-measured record are retained. Changing $P$ at fixed measurement therefore changes the accessible tangent geometry without changing the underlying quantum state or POVM.

\section{Rank baseline and spectral orientation}

\subsection{Random relative orientation}

To define a rank-only reference, keep $C$ fixed and randomize only the orientation of a rank-$r$ projector $P$. For a Haar-uniform point on the real Grassmann manifold $\mathrm{Gr}(r,N)$, rotational invariance gives
\begin{equation}
\mathbb{E}_P P=\frac{r}{N}I,
\end{equation}
and therefore
\begin{equation}
\mathbb{E}_P R(P,C)=\frac{r}{N},
\qquad
\mathbb{E}_P\rho(P,C)=1.
\label{eq:rank-baseline}
\end{equation}
The identity is a standard random-subspace result \cite{CollinsMatsumoto2017,Bendokat2024,Fernandez2025}. In the isotropic setting, the same mean $r/N$ appears as the exact readout-rank law for a fixed measurement record \cite{AitHaddou2026IsotropicRankLaws}; here we reinterpret it as a random-relative-orientation baseline for an arbitrary trace-one tangent covariance. Its role here is to provide a controlled null model for readout orientation inside a fixed measurement-induced score space.

Using standard Grassmann projector moments \cite{CollinsMatsumoto2017}, we obtain
\begin{equation}
\mathrm{Var}_P(\rho\mid C)
=\frac{2(N-r)\left[N\Tr(C^2)-1\right]}
{r(N-1)(N+2)}.
\label{eq:exact-var}
\end{equation}
Using $N-r\le N-1$ and $N\Tr(C^2)-1\le(N+2)\Tr(C^2)$ yields
\begin{equation}
\mathrm{Var}_P(\rho\mid C)
\le \frac{2\Tr(C^2)}{r}
=\frac{2}{r\deff(C)}.
\label{eq:var-bound}
\end{equation}
Consequently,
\begin{equation}
\Pr\left(|\rho-1|\ge\epsilon\mid C\right)
\le
\min\left\{1,\frac{2}{\epsilon^2r\deff(C)}\right\}.
\label{eq:cheb}
\end{equation}

Equation~\eqref{eq:var-bound} is the key separation used below. A full projector-moment derivation is given in Appendix~\ref{app:grassmann}. Rank-typical overlap can coexist with strong anisotropy because concentration depends on $r\deff$, not on $\deff/N$ approaching one. Conversely, a physical readout can lie far above or below the rank reference if its orientation is systematically correlated or anticorrelated with the leading eigenspaces of $C$.

\subsection{Fixed-weight diagonal readout}

For computational-basis measurement on $n$ qubits with full support, the centered score-space dimension is
\begin{equation}
N_n=2^n-1.
\end{equation}
The span of diagonal Pauli-$Z$ strings through fixed weight $k$ has rank
\begin{equation}
r_k(n)=\sum_{j=1}^{k}\binom{n}{j}
=\frac{n^k}{k!}\left[1+O(n^{-1})\right].
\label{eq:rk}
\end{equation}
Under random relative orientation, the expected retained fraction is therefore
\begin{equation}
\mathbb{E}R_k=\frac{r_k(n)}{2^n-1}
=O\!\left(\frac{n^k}{2^n}\right).
\label{eq:fixed-weight}
\end{equation}
This is a baseline for comparison, not a statement that any physical ansatz approaches a Haar-random orientation as $n$ increases. The full- and fixed-charge rank bookkeeping is collected in Appendix~\ref{app:walsh_rank}.

\begin{figure}[t]
\centering
\includegraphics[width=.98\linewidth]{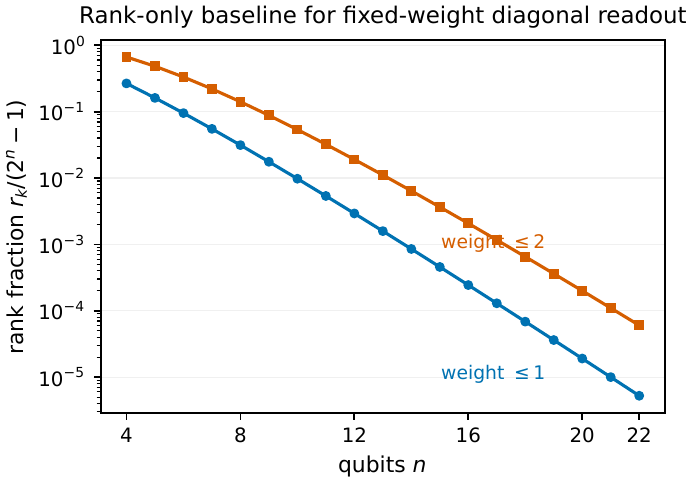}
\caption{Rank-only scale for fixed-weight diagonal readout under full computational-basis support. For fixed Pauli weight, the retained subspace rank grows polynomially while the centered score-space dimension grows exponentially. The plotted fractions are geometric baselines only; physical retention depends on spectral orientation.}
\label{fig:rank-fraction-production}
\end{figure}

\subsection{Optimal and cross-fitted same-rank subspaces}

For a fixed covariance $C$ with eigenvalues $\lambda_1\ge\cdots\ge\lambda_N$, the Ky Fan variational principle \cite{KyFan1949} gives
\begin{equation}
\max_{\operatorname{rank}(P)=r}\Tr(PC)=\sum_{j=1}^{r}\lambda_j.
\label{eq:kyfan}
\end{equation}
This optimization is standard linear algebra; its use here is diagnostic. It answers whether poor physical retention is forced by rank or caused by orientation. Because the same-sample optimum is statistically optimistic, the operational comparison uses a cross-fitted projector (Appendix~\ref{app:crossfit}): the leading rank-$r$ score subspace is estimated from an independent set of tangent directions and evaluated on held-out tangents. We compare this aligned cross-fit subspace with both the physical low-weight readout and a random rank-matched projector.

\section{Numerical strategy}

The numerical campaign treats a circuit instance, not an individual tangent direction, as the independent statistical unit. Within each circuit, tangent directions estimate the score covariance and readout statistics. Confidence intervals are obtained by nonparametric bootstrap over independent circuit instances \cite{Efron1979}. The generic aggregate is family-balanced so that no ansatz family dominates by sample count. The source code, frozen experiment profiles, aggregate tables, shard-level outputs, and paper-facing result summaries used for the numerical claims are archived in the public reproducibility repository \cite{AitHaddou2026MeasurementAccessible}.

The baseline campaign uses computational-basis measurement, depth proportional to system size, and diagonal $Z$ readouts through weight one and two. The nonconserving ensemble includes RY--RZ--CZ, SU2--CNOT, SU2--CZ, random-matching SU2--CZ, and Haar-$U(4)$ brickwork families. A half-filled number-conserving RZ--XY family supplies the structured $U(1)$ case study. Larger-size extensions and dedicated cross-fit experiments use frozen profiles documented in the repository \cite{AitHaddou2026MeasurementAccessible}.

For the operational comparison we hold fixed the circuit, the full computational-basis record, the readout rank, and the evaluation shot budget. We compare three rank-matched readouts: (i) the physical low-weight span, (ii) an independent random subspace, and (iii) a cross-fitted aligned subspace learned from independent tangent data. For an evaluation tangent $v$, the retained score energy $\|P u_v\|^2$ controls the directional signal available in that readout. Multiplying by the full-record Fisher scale gives the raw directional gradient-energy proxy used in the experiment. Finite-shot signal-to-noise is computed under the multinomial shot model at a fixed budget of $10^4$ shots. The independent alignment sample used to estimate the cross-fitted projector is a separate calibration resource and is not counted as free measurement cost in this fixed-evaluation-budget comparison.

\begin{figure*}[t]
\centering
\includegraphics[width=.49\textwidth]{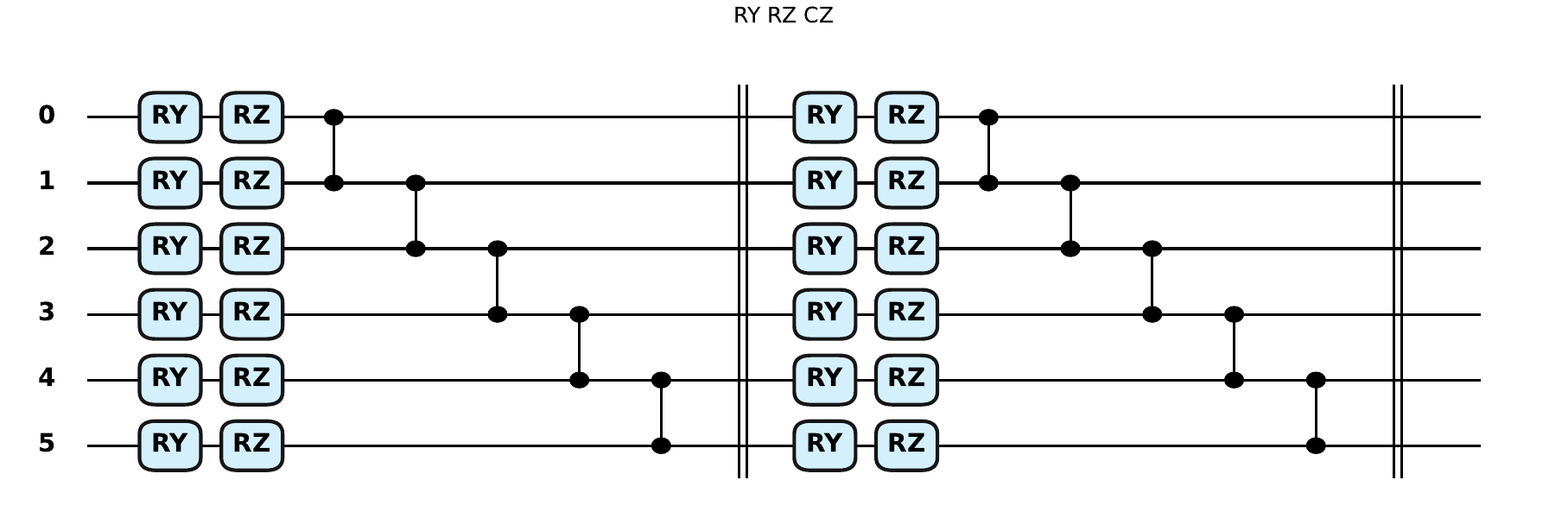}\hfill
\includegraphics[width=.49\textwidth]{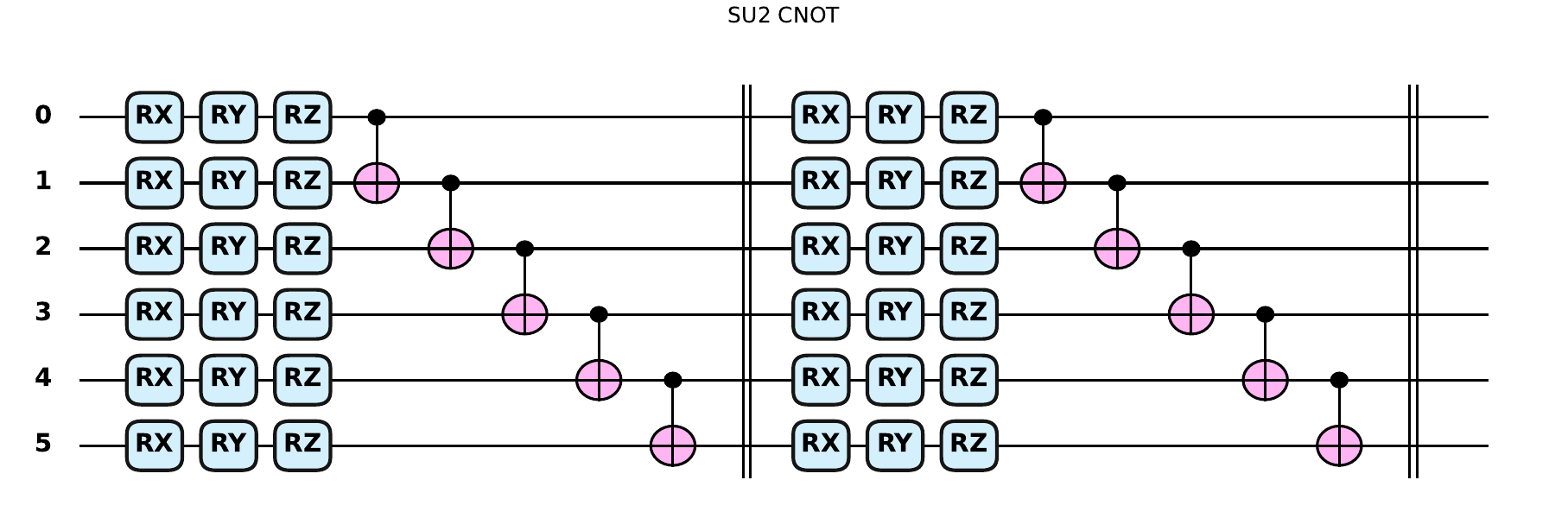}\\[2pt]
\includegraphics[width=.49\textwidth]{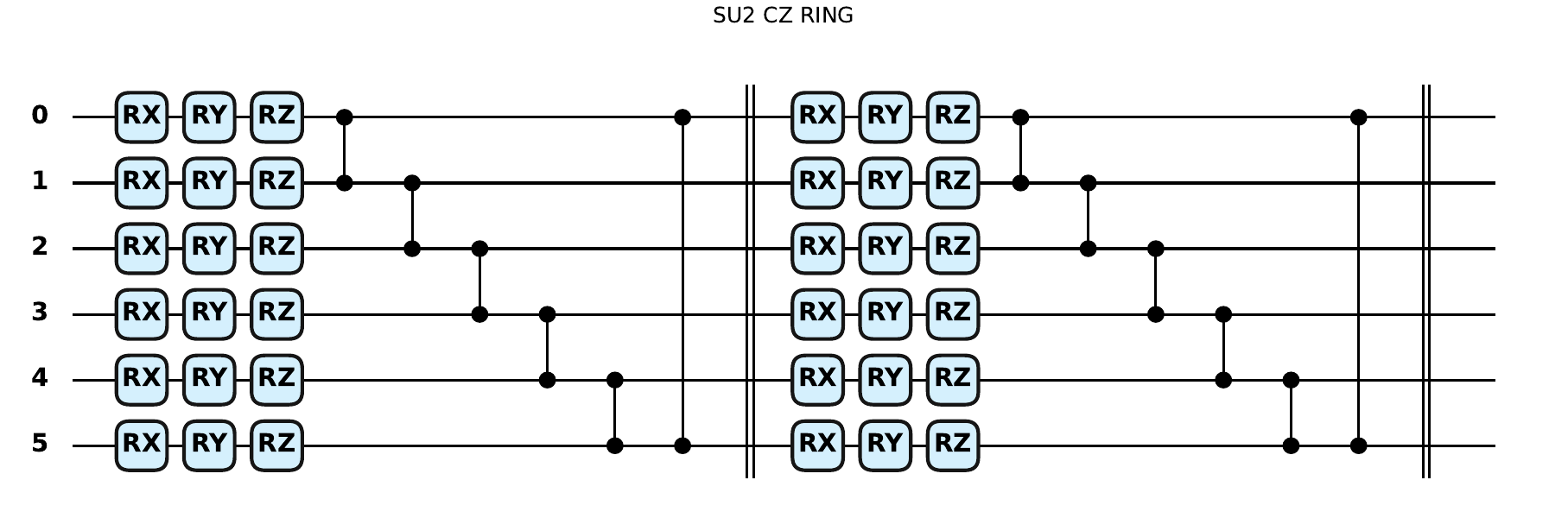}\hfill
\includegraphics[width=.49\textwidth]{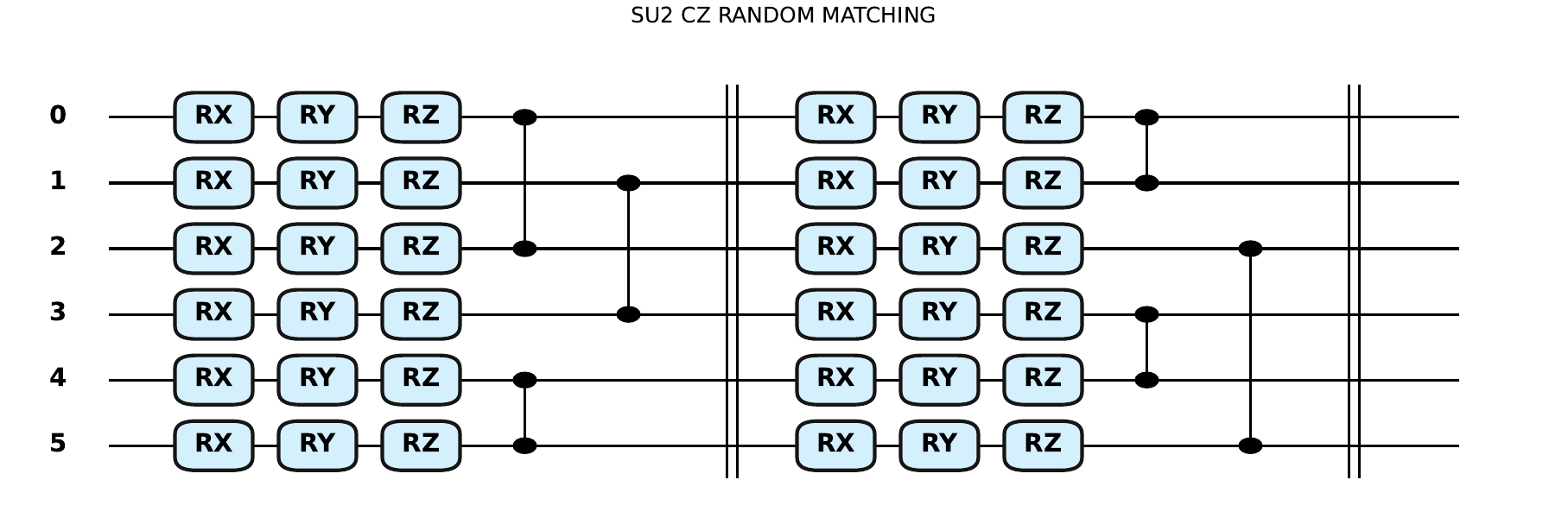}\\[2pt]
\includegraphics[width=.49\textwidth]{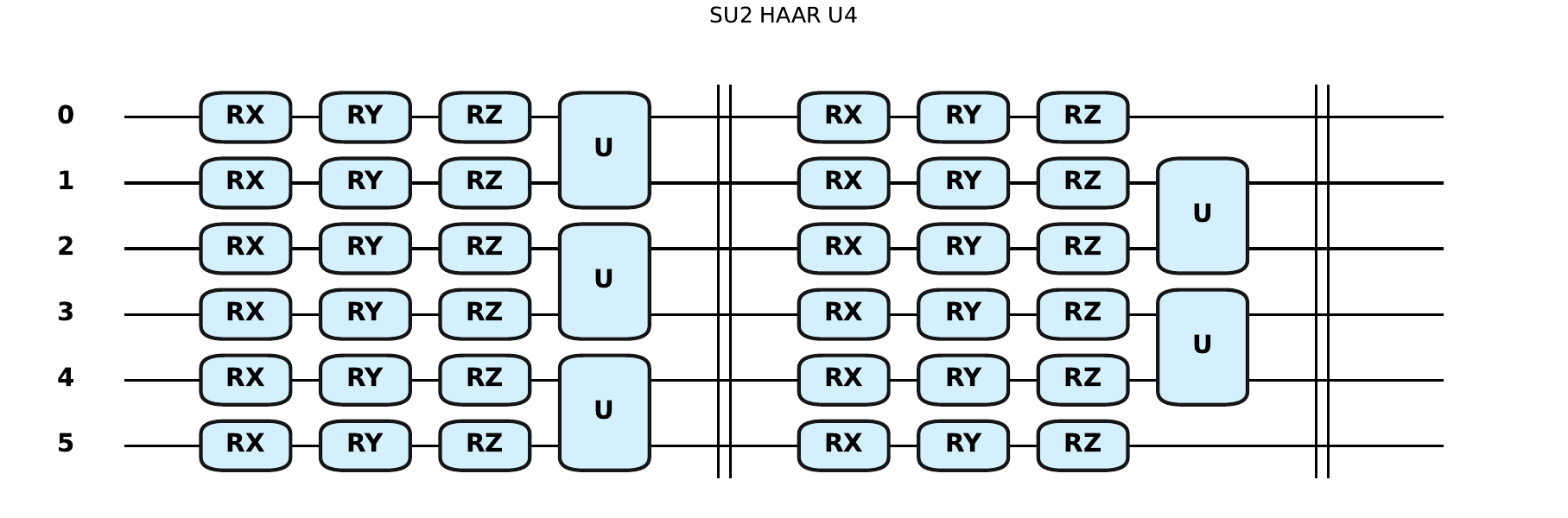}\hfill
\includegraphics[width=.49\textwidth]{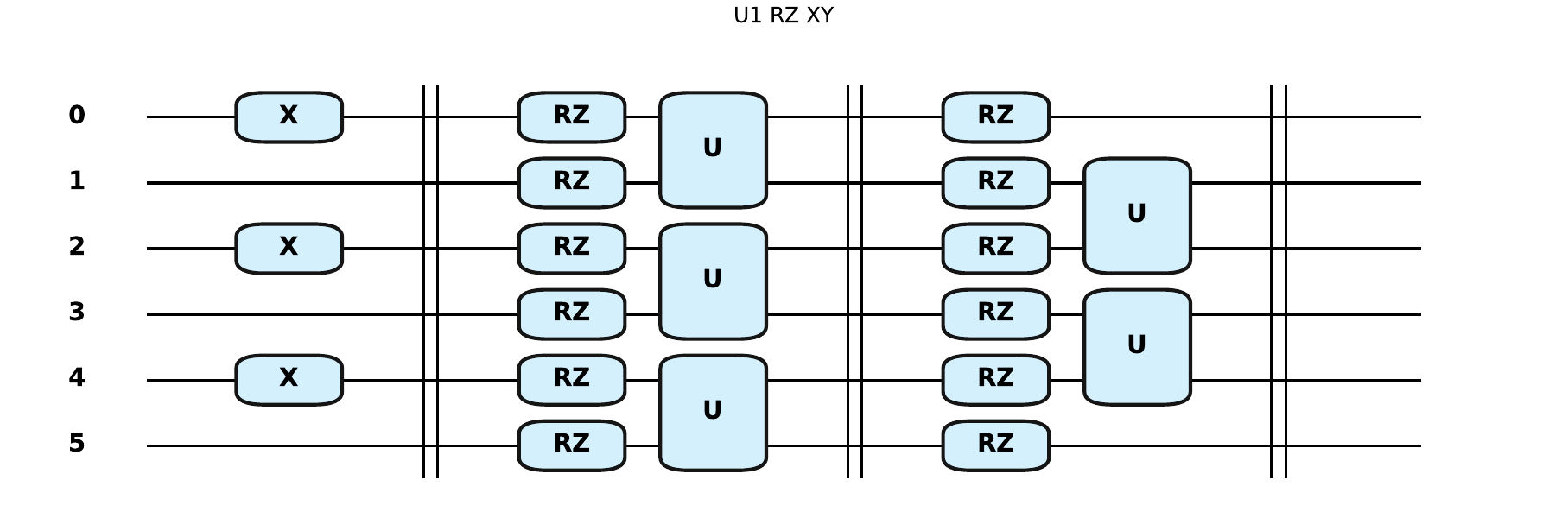}
\caption{Circuit families used in the numerical campaign, rendered directly from PennyLane circuit definitions \cite{Bergholm2018}. From top left to bottom right: RY--RZ--CZ line, SU2--CNOT line, SU2--CZ ring, SU2--CZ random matching, SU2--Haar-$U(4)$ brickwork, and the half-filled $U(1)$ RZ--XY line. For legibility the drawings show $n=6$ and two representative layers; the production simulations use the depths and system sizes specified in the numerical protocol. The $U(1)$ panel begins from the alternating half-filled computational-basis state, matching the simulation code. Generic unitary boxes in the last row represent the actual two-qubit matrices used by the simulator rather than a gate decomposition.}
\label{fig:circuits}
\end{figure*}

\section{Rank typicality without isotropy in generic circuits}

For the family-balanced generic ensemble, the one- and two-body readouts remain within the preregistered rank-equivalence band through $n=16$. The targeted Haar-$U(4)$ stress test at $n=18$ gives
\begin{equation}
\rho_1=0.929\;[0.910,0.950],
\qquad
\rho_2=0.958\;[0.952,0.966].
\label{eq:n18-rho}
\end{equation}
The $n=18$ point is a single-architecture stress test rather than the family-balanced estimator used through $n=16$ \cite{AitHaddou2026MeasurementAccessible}.

These near-rank values do not indicate isotropy. The normalized purity $N\Tr(C^2)$ grows from approximately $1.35$ at $n=6$ to $6.17$ at $n=12$, $17.52$ at $n=14$, and $49.84$ at $n=16$; the Haar-$U(4)$ $n=18$ stress point is approximately $146.8$. Correspondingly, $\deff/N$ falls from $0.746$ at $n=6$ to $0.1767$ at $n=12$ and $0.0233$ at $n=16$, with the $n=18$ Haar-$U(4)$ value near $0.00683$. Thus the physical low-weight readout can remain close to the rank reference while the covariance becomes highly concentrated spectrally \cite{AitHaddou2026MeasurementAccessible}.

\begin{figure*}[t]
\includegraphics[width=\textwidth]{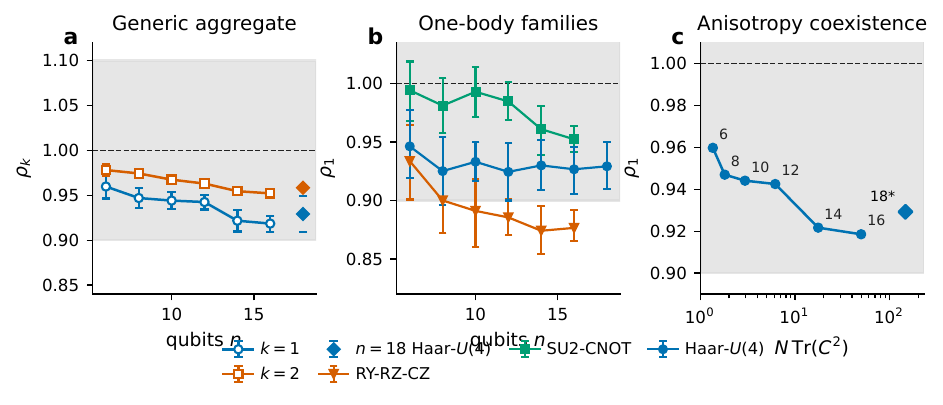}
\caption{Rank-normalized retention and covariance anisotropy. Family-balanced generic low-weight readouts remain close to the rank reference over the tested sizes even while $N\Tr(C^2)$ grows and $\deff/N$ decreases. Architecture-resolved data show systematic deviations, demonstrating that the family-balanced result is not a claim of literal Haar orientation.}
\label{fig:rank}
\end{figure*}

Architecture-resolved results make the orientation dependence explicit. The RY--RZ--CZ family develops a persistent one-body deficit, with mean $\rho_1\simeq0.874$ at $n=14$ and $0.877$ at $n=16$. The full computational-basis Fisher fraction remains order one, so this deficit is specific to the low-weight projection rather than to disappearance of the measured tangent signal itself. We do not assign a microscopic gate-level cause in the present work \cite{AitHaddou2026MeasurementAccessible}.

\section{Equal rank does not imply equal accessibility}

The rank baseline becomes physically informative only if equal-rank subspaces can produce measurably different outcomes. The spectral experiments test this directly.

For Haar-$U(4)$ at $n=12$, the physical one-body readout retains
\begin{equation}
R_{\rm phys}=2.71\times10^{-3}.
\end{equation}
A rank-matched cross-fitted leading tangent subspace retains
\begin{equation}
R_{\rm xfit}=3.06\times10^{-2},
\end{equation}
while the same-sample Ky Fan optimum retains $9.15\times10^{-2}$. The physical readout therefore misses tangent mass that is available at exactly the same rank. This is the central distinction between rank and orientation \cite{AitHaddou2026MeasurementAccessible}.

The $U(1)$ circuit behaves differently. At $n=12$, the physical one-body readout retains $0.292$, the cross-fitted aligned subspace $0.375$, and the same-sample Ky Fan subspace $0.444$. The physical readout is not optimal, but it is already much closer to the leading tangent subspace than the generic Haar-$U(4)$ readout \cite{AitHaddou2026MeasurementAccessible}.

\begin{figure*}[t]
\includegraphics[width=\textwidth]{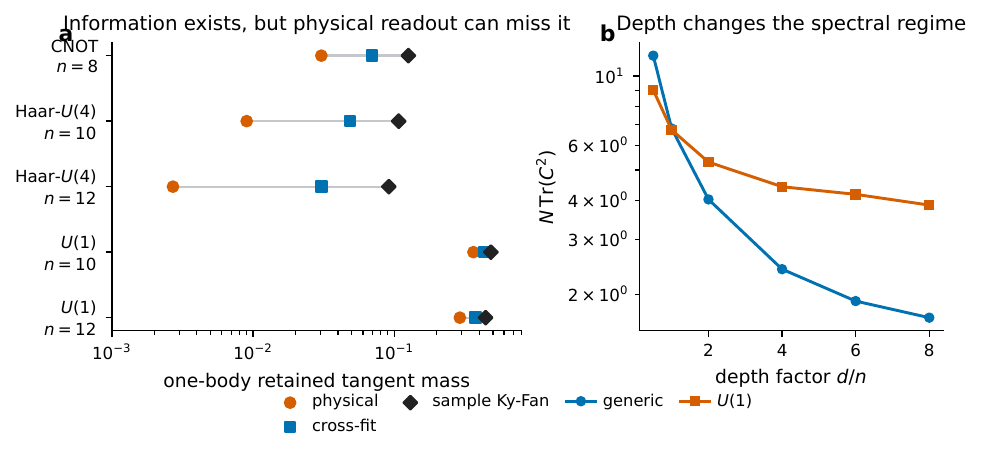}
\caption{Same-rank spectral comparison. Physical low-weight, cross-fitted aligned, and same-sample Ky Fan subspaces have identical rank but can retain very different tangent mass. The orientation gap is large in generic Haar-$U(4)$ circuits and substantially smaller in the $U(1)$ family.}
\label{fig:spectrum}
\end{figure*}

\section{Operational consequence at fixed rank and shot budget}

The same-rank comparison is not only geometric. We evaluate directional gradient energy and finite-shot signal-to-noise while holding the circuit, computational-basis record, readout rank, and shot budget fixed.

For Haar-$U(4)$ circuits, replacing the physical one-body readout by the cross-fitted aligned subspace increases the mean directional gradient-energy proxy by factors
\begin{equation}
2.963,\quad 4.639,\quad 9.584
\end{equation}
for $n=8,10,12$, respectively. The corresponding finite-shot SNR gains are
\begin{equation}
1.734,\quad 2.125,\quad 3.111.
\end{equation}
A random rank-matched subspace remains close to the physical generic baseline, as expected from the rank-only null model \cite{AitHaddou2026MeasurementAccessible}.

The $U(1)$ family shows a smaller orientation gain because its physical readout is already aligned. At $n=8,10,12$, the aligned-to-physical gradient-energy ratios are approximately $1.148$, $1.189$, and $1.210$, with SNR gains $1.068$, $1.097$, and $1.112$. The contrast between the two families is itself informative: equal rank leaves substantial room for readout design in generic circuits, while the structured $U(1)$ readout already captures much of the accessible leading tangent structure \cite{AitHaddou2026MeasurementAccessible}.

These quantities are not supervised-loss gradient variances. Their precise directional definitions and fixed-shot interpretation are summarized in Appendix~\ref{app:operational}. They are controlled directional-signal diagnostics under a fixed measurement and finite-shot model. Their purpose is to show that spectral orientation has an operational consequence without introducing labels, a classical head, or an optimization trajectory.

\section{Structured case study: a half-filled $U(1)$ circuit}

The number-conserving RZ--XY family provides a regime in which physical low-weight diagonal readout lies far above the full-space random-orientation baseline. In the original large-size campaign, the one-body retained fraction remains of order $10^{-1}$ while the corresponding generic/Haar rank baseline becomes exponentially small in $n$. At $n=18$, for example, the one-body $U(1)$ retention is approximately $0.215$, compared with the Haar-$U(4)$ value $6.38\times10^{-5}$; the two-body values are approximately $0.401$ and $6.25\times10^{-4}$, respectively \cite{AitHaddou2026MeasurementAccessible}.

\begin{figure*}[t]
\centering
\includegraphics[width=.95\textwidth]{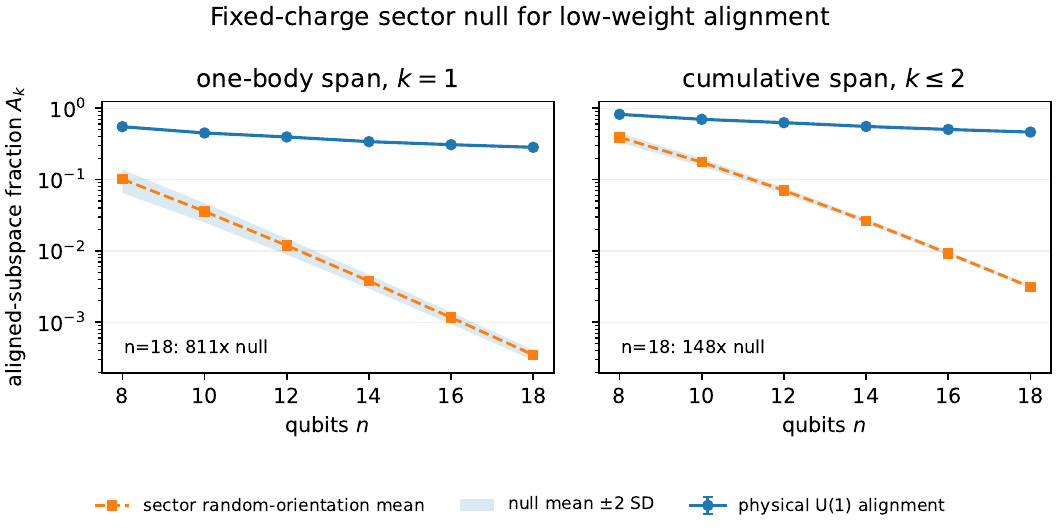}
\caption{Sector-corrected random-orientation control for the half-filled $U(1)$ alignment statistic. The null draws the leading rank-$r_1$ subspace randomly inside the same half-filled centered score sector, rather than comparing with the full computational-basis score space. Points are the measured cross-fitted alignment means with circuit-bootstrap intervals; dashed curves are the exact random-orientation means and shaded bands show two null standard deviations from Grassmann projector moments. The observed alignment remains orders of magnitude above the sector null at the largest sizes.}
\label{fig:u1-sector-null}
\end{figure*}

The stronger diagnostic is direct subspace overlap. Let $P_{\rm top}$ denote the cross-fitted leading tangent subspace of rank equal to the centered one-body readout rank, and let $P_{\le k}$ denote the cumulative low-weight Walsh span. We use
\begin{equation}
A_k=\frac{1}{r_1}\Tr(P_{\le k}P_{\rm top})
\label{eq:Ak}
\end{equation}
to quantify how much of the leading tangent subspace lies in low-weight sectors. A sector-corrected random-orientation control is available without assuming full-space isotropy. If $P_{\rm top}$ is Haar-random inside the same half-filled centered score sector, then $\mathbb E A_k=r_{\le k}/N_{\rm hf}$; its exact variance follows from Eq.~\eqref{eq:exact-var} by setting $C=P_{\le k}/r_{\le k}$ and the random projector rank to $r_1$. At $n=18$ the corresponding null means are $0.00035$ for $k=1$ and $0.00313$ for $k\le2$, while the observed overlaps are approximately $811\times$ and $148\times$ larger, respectively. Thus the persistent alignment is not an artifact of comparing a fixed-charge family with the full $2^n$ score-space rank scale. Figure~\ref{fig:u1-sector-null} summarizes the sector-corrected comparison over the full tested window. This is a VQC-specific use of a standard projector/subspace-overlap construction rather than a claim that subspace overlap itself is new \cite{Fernandez2025}. Across $n=8,10,12,14,16,18$, the one-body values are $0.552$, $0.450$, $0.396$, $0.341$, $0.309$, and $0.284$, respectively. The final $n=18$ estimate is
\begin{equation}
A_1(18)=0.283586\;[0.277044,0.290865],
\end{equation}
while the cumulative weight-through-two overlap is
\begin{equation}
A_{\le2}(18)=0.463538\;[0.457425,0.469591].
\end{equation}
Each size uses 20 independent circuit instances \cite{AitHaddou2026MeasurementAccessible}.

Over the tested window $n=8$--$18$, two-parameter log-space model comparison favors a finite-size power form over an exponential form. The fitted exponents are $0.8223\,[0.7694,0.8735]$ for $A_1$ and $0.7041\,[0.6798,0.7281]$ for $A_{\le2}$, with $\Delta\mathrm{AICc}=14.39$ and $12.82$, respectively, in favor of the power model. These are finite-size model-discrimination results only; they are not asymptotic lower bounds and are not identified with a hydrodynamic exponent \cite{AitHaddou2026MeasurementAccessible}.

\begin{figure*}[t]
\centering
\includegraphics[width=.93\textwidth]{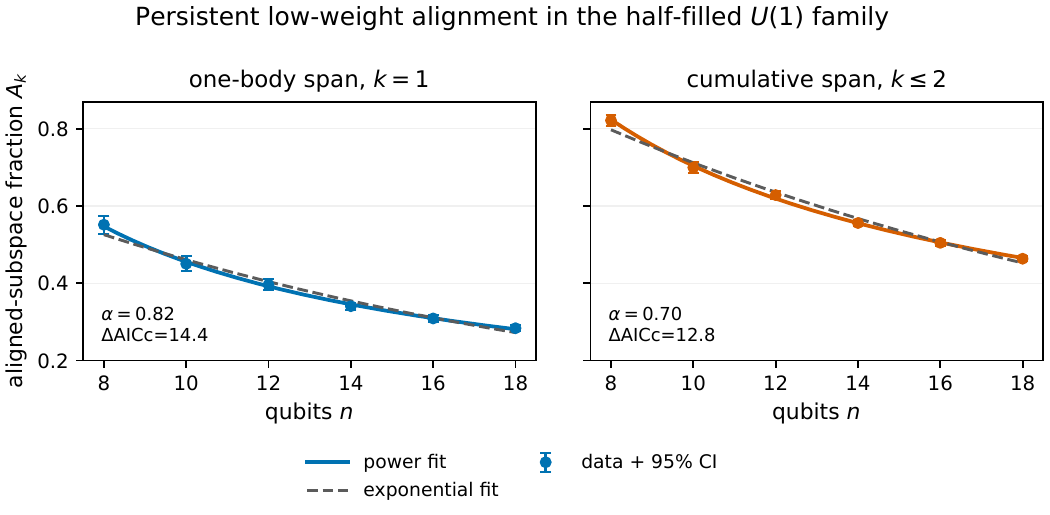}
\caption{Finite-size low-weight alignment in the half-filled $U(1)$ family. Points and error bars are circuit-level means and 95\% bootstrap intervals for the cross-fitted leading rank-$r_1$ tangent subspace. Solid curves are the two-parameter power fits and dashed curves are exponential fits over $n=8$--$18$. Positive $\Delta$AICc favors the power model over the tested window. The comparison is descriptive finite-size model selection, not an asymptotic or hydrodynamic theorem.}
\label{fig:u1-scaling}
\end{figure*}

This observation should not be read as a new theorem that $U(1)$ symmetry improves trainability. Hamming-weight-preserving trainability, Lie-algebraic restrictions, low-bodyness concentration in structured quantum models, and readout-visible sector coherence in noisy equivariant circuits are established neighboring phenomena \cite{Monbroussou2025,Fontana2024,Ragone2024,Bermejo2026,UgailHoward2026}. In addition, fixed-charge slice geometry supplies a non-dynamical structural null \cite{Filmus2016}, while conserved-density hydrodynamics supplies a distinct dynamical mechanism in random circuits \cite{Khemani2018,Rakovszky2018}. The present result is narrower: in this specific half-filled variational family, the leading measurement-induced tangent subspace is unusually visible to low-weight diagonal $Z$ readout under the rank-controlled metric used throughout this paper. The same family was already identified in our isotropic-rank analysis as a strong violation of the support-corrected rank law \cite{AitHaddou2026IsotropicRankLaws}; the present analysis resolves that deviation spectrally in terms of persistent low-weight alignment.

A separately archived pilot tests sensitivity to exact conservation at fixed trainable parameter count. At $n=8$, the unperturbed one-body alignment is $A_1=0.517\,[0.486,0.549]$. A symmetry-preserving $R_Z$ perturbation of strength $\epsilon=0.3$ leaves $A_1=0.481\,[0.454,0.518]$, whereas a nontrainable symmetry-breaking $R_X$ perturbation at the same strength gives $A_1=0.0295\,[0.0248,0.0338]$. The physical one-body retention simultaneously falls from $0.430$ to $0.0325$ in the breaking control \cite{AitHaddou2026MeasurementAccessible}. We use this as a sensitivity control, not as a mechanism-identification experiment: it does not separate readout--charge compatibility, fixed-charge harmonic geometry, restricted controllability, and hydrodynamic slow modes.

\begin{figure}[t]
\centering
\includegraphics[width=.98\linewidth]{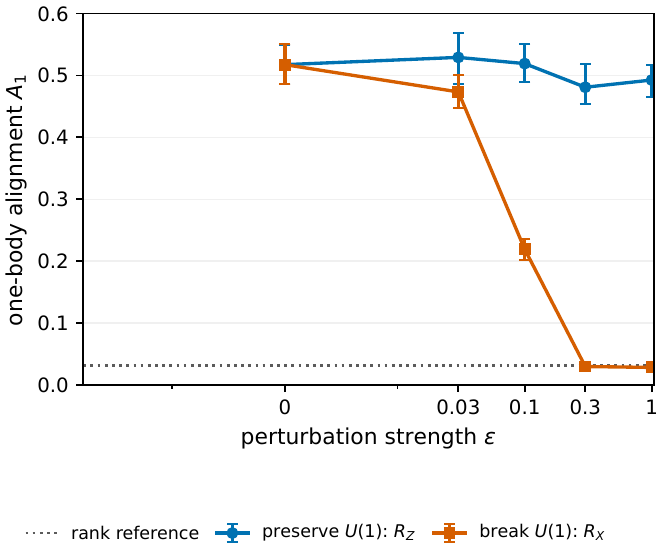}
\caption{Verified symmetry-breaking pilot at $n=8$. Symmetry-preserving $R_Z$ perturbations leave the one-body tangent-subspace overlap comparatively stable, while nontrainable $R_X$ perturbations that break charge conservation suppress it toward the full-score-space rank scale. Error bars are 95\% circuit-bootstrap intervals over 10 circuit instances. The dotted line is a visual full-space rank reference after charge breaking, not a fit and not a sector-Haar null.}
\label{fig:symmetry-breaking}
\end{figure}

\section{Full measurement record versus restricted readout}

The complete computational-basis record retains a similar order-one fraction of QFI in generic and $U(1)$ circuits. Across the generic data, $F_{\rm full}/F_Q$ is approximately $0.49$, while the $U(1)$ values remain near $0.47$--$0.48$ \cite{AitHaddou2026MeasurementAccessible}. The orders-of-magnitude separation between the two circuit classes therefore appears primarily after projection onto the low-weight readout span.

This comparison prevents a misleading interpretation of the $U(1)$ result. The structured circuit does not simply preserve vastly more classical Fisher information under computational-basis measurement. Rather, a much larger fraction of the measured tangent information is oriented toward the low-weight functions retained by the readout.

\section{Discussion}

The results support a three-part separation. Rank fixes a null accessibility scale. The covariance spectrum controls how broadly random rank-matched subspaces fluctuate around that scale. The physical orientation of the readout relative to the tangent eigenspaces controls the actual retained mass. These quantities can vary independently enough that rank-typical retention coexists with strong anisotropy, while equal-rank readouts yield substantially different finite-shot signal.

The strongest evidence for the usefulness of this separation is the controlled same-rank experiment. The circuit and quantum measurement are unchanged; the rank and shot budget are unchanged; only the retained score subspace is changed. In generic Haar-$U(4)$ circuits, cross-fitted alignment increases the directional gradient-energy proxy by nearly an order of magnitude at $n=12$. This effect cannot be attributed to adding more observables, increasing the measurement budget, or changing the underlying state family. It is an orientation effect inside the same measurement-induced score space.

The random-orientation formulas should therefore be interpreted as a baseline, not as the novelty claim by themselves. Similar Grassmann overlap identities are standard in random-projection theory and appear in neighboring classical subspace-overlap problems \cite{Fernandez2025}. Random-measurement work also establishes a distinct Haar-averaged CFIM--QFIM relation with variance and concentration bounds when the measurement basis itself is randomized \cite{LuSha2025}. Likewise, the Ky Fan optimum is standard spectral optimization \cite{KyFan1949}. The contribution here is the rank-controlled organization of these ingredients around measurement-induced VQC tangent scores, together with physical, random, and cross-fitted equal-rank controls and a finite-shot operational test.

The $U(1)$ family illustrates why this organization is useful. Its physical readout is far from a generic random orientation and already points toward leading tangent directions. Existing Hamming-weight-preserving trainability and Lie-algebraic theories \cite{Monbroussou2025,Fontana2024,Ragone2024}, readout-visible equivariant coherence \cite{UgailHoward2026}, fixed-charge slice geometry \cite{Filmus2016}, and conserved-density hydrodynamics \cite{Khemani2018,Rakovszky2018} offer plausible but non-equivalent explanations for this structure. Determining which mechanism controls the observed overlap requires additional interventions---for example varying locality independently of conservation, comparing against fixed-charge random baselines, or resolving depth scaling---and is left for separate work.

Finally, measurement accessibility is not the same as task usefulness. A supervised learning problem introduces another layer: the task gradient may occupy only a subset of the accessible tangent geometry. The present framework therefore sits between state-space geometry and task-level optimization geometry. It quantifies what the chosen measurement and readout make available before asking whether a specific loss uses those directions.

\section{Conclusion}

We studied the fraction of measurement-induced tangent geometry retained by a restricted readout. For a trace-one tangent-score covariance $C$ and rank-$r$ readout projector $P$, the retained mass $R=\Tr(PC)$ makes the role of readout orientation explicit. The standard Grassmann mean $r/N$ provides a rank-only reference, while the spectrum of $C$ sets the null width through $\deff$.

The numerical results show why rank and orientation should be separated. Generic low-weight readouts can be close to their rank baseline even when the covariance is strongly anisotropic. At the same rank, cross-fitted leading tangent subspaces can retain much more information than physical low-weight observables, and that difference produces larger directional signal and finite-shot SNR at fixed measurement resources. A half-filled $U(1)$ circuit provides the complementary structured regime in which the physical low-weight readout is already strongly aligned with the leading tangent subspace.

The practical conclusion is therefore not that low rank is intrinsically bad, nor that symmetry is intrinsically good. It is that readout rank specifies only how many score directions are retained. Which directions are retained---their spectral orientation relative to the tangent covariance---controls how much measured tangent information remains accessible.

\clearpage

\appendix

\section{Score-space construction and Fisher normalization}
\label{app:score_geometry}

For completeness, we collect here the conventions that connect the fixed measurement record to the normalized score geometry used in the main text.  Let $p_\theta(x)$ be the probability of outcome $x$ under the fixed computational-basis measurement and let $v$ be a parameter-space direction.  The directional probability tangent is $\partial_v p_\theta(x)$.  On the regular support, the Fisher-weighted tangent may be written as
\begin{equation}
 s_v(x)=\frac{\partial_v p_\theta(x)}{\sqrt{p_\theta(x)}}.
\label{eq:app-score}
\end{equation}
The probability-normalization constraint removes the constant mode, leaving an $N$-dimensional centered real score space.  The numerical protocol normalizes every regular tangent direction according to
\begin{equation}
 u_v=\frac{s_v}{\|s_v\|_2},\qquad \|u_v\|_2=1,
\label{eq:app-normalized-score}
\end{equation}
and forms
\begin{equation}
 C=\mathbb E_v[u_vu_v^{\mathsf T}],\qquad \Tr C=1.
\label{eq:app-C}
\end{equation}
Before the unit normalization in Eq.~\eqref{eq:app-normalized-score}, $\|s_v\|_2^2$ is the classical Fisher information carried by the fixed measurement along $v$ \cite{BraunsteinCaves1994,LuLuoOh2012}.  The normalized covariance $C$ therefore describes how the \emph{directions} of measurement-visible tangent variation are distributed after the overall directional Fisher scale has been factored out.

For an orthogonal readout projector $P$, linearity and cyclicity of the trace give
\begin{align}
\mathbb E_v\|Pu_v\|_2^2
&=\mathbb E_v\,u_v^{\mathsf T}Pu_v \\
&=\Tr\!\left(P\,\mathbb E_v[u_vu_v^{\mathsf T}]\right)
=\Tr(PC),
\label{eq:app-R}
\end{align}
which is the operational identity used in Eq.~\eqref{eq:R-operational}.

\section{Grassmann random-orientation moments}
\label{app:grassmann}

We give the complete moment calculation underlying Eqs.~\eqref{eq:rank-baseline}--\eqref{eq:var-bound}.  Let $P$ be Haar-uniform on the real Grassmann manifold $\mathrm{Gr}(r,N)$.  Orthogonal invariance implies
\begin{equation}
\mathbb E P=\frac rN I.
\end{equation}
Hence, for any fixed trace-one $C$,
\begin{equation}
\mathbb E\Tr(PC)=\Tr\!\left(C\,\mathbb EP\right)=\frac rN.
\end{equation}
The second moment of a random orthogonal projector has the invariant tensor form
\begin{equation}
\mathbb E[P_{ij}P_{kl}]
=a\,\delta_{ij}\delta_{kl}
+b\,(\delta_{ik}\delta_{jl}+\delta_{il}\delta_{jk}),
\label{eq:app-projector-second}
\end{equation}
with coefficients fixed by $P^2=P$ and $\Tr P=r$ \cite{CollinsMatsumoto2017}.  Solving these constraints and contracting Eq.~\eqref{eq:app-projector-second} with $C_{ji}C_{lk}$ yields
\begin{equation}
\operatorname{Var}[\Tr(PC)]
=\frac{2r(N-r)}{N^2(N-1)(N+2)}
\left[N\Tr(C^2)-1\right].
\label{eq:app-var-R}
\end{equation}
Since $\rho=N\Tr(PC)/r$, Eq.~\eqref{eq:app-var-R} gives
\begin{equation}
\operatorname{Var}(\rho)
=\frac{2(N-r)[N\Tr(C^2)-1]}
{r(N-1)(N+2)}.
\label{eq:app-var-rho}
\end{equation}
Using $N-r\le N-1$ and $N\Tr(C^2)-1\le (N+2)\Tr(C^2)$ gives
\begin{equation}
\operatorname{Var}(\rho)\le \frac{2}{r d_{\rm eff}},
\qquad d_{\rm eff}=\frac1{\Tr(C^2)}.
\label{eq:app-deff-bound}
\end{equation}
Thus concentration depends on the product $r d_{\rm eff}$, not on $d_{\rm eff}/N$ approaching unity.

\begin{figure}[t]
\centering
\includegraphics[width=.96\linewidth]{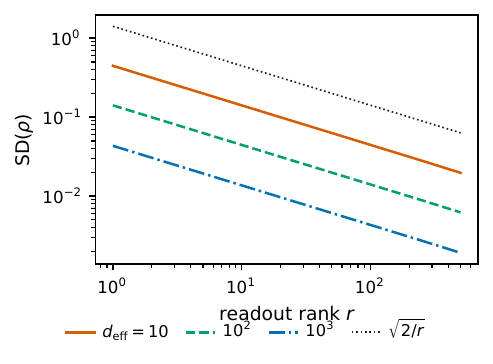}
\caption{Exact random-orientation fluctuation scale for representative effective dimensions.  The standard deviation of $\rho$ narrows with increasing readout rank and effective spectral dimension; this is compatible with a strongly anisotropic covariance whenever $r d_{\rm eff}$ is large.}
\label{fig:app-null-width}
\end{figure}

\section{Fixed-weight diagonal readout spaces}
\label{app:walsh_rank}

For full computational-basis support, the centered outcome space has dimension $N=2^n-1$.  Distinct nonidentity diagonal Pauli strings are orthogonal in the uniform Walsh basis, so the cumulative weight-through-$k$ span has dimension
\begin{equation}
 r_{\le k}=\sum_{j=1}^k\binom nj.
\label{eq:app-rank-full}
\end{equation}
For fixed $k$ and large $n$, $r_{\le k}=n^k/k!\,[1+O(n^{-1})]$, which gives the exponentially small full-space rank fraction used in the main text.

In the half-filled fixed-charge sector, the outcome support contains $\binom{n}{n/2}$ basis strings and the centered score dimension is
\begin{equation}
 N_{\rm hf}=\binom{n}{n/2}-1.
\end{equation}
The fixed-charge identity $\sum_i Z_i=0$ on the half-filled support removes one one-body degree of freedom, giving
\begin{equation}
 r_1=n-1.
\end{equation}
For the cumulative one- and two-body diagonal span, the corresponding centered dimension used in the numerical protocol is
\begin{equation}
 r_{\le2}=\binom n2-1.
\end{equation}
These sector-corrected ranks are geometric bookkeeping identities; they do not by themselves imply any particular physical orientation of the tangent covariance.  Fixed-charge harmonic analysis provides a broader mathematical context for low-degree functions on a slice \cite{Filmus2016}.

\begin{figure}[t]
\centering
\includegraphics[width=.96\linewidth]{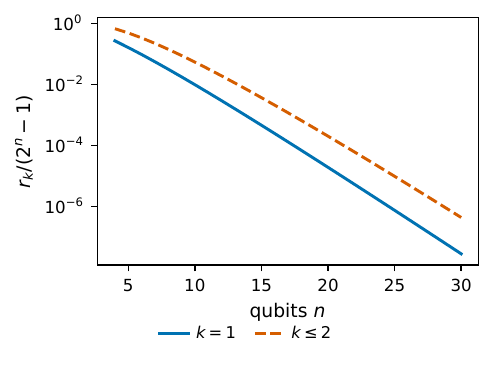}
\caption{Rank fractions for low-weight diagonal readouts under full computational-basis support.  Polynomially growing low-weight spaces occupy an exponentially shrinking fraction of the centered record.}
\label{fig:app-rank-fractions}
\end{figure}

\section{Cross-fitting, Ky Fan benchmark, and statistical unit}
\label{app:crossfit}

For a covariance estimate $\widehat C_{\rm fit}$ obtained from one tangent ensemble, let $Q_r$ contain its leading $r$ eigenvectors and define the aligned projector $P_{\rm xfit}=Q_rQ_r^{\mathsf T}$.  Retention is then evaluated on an independent tangent ensemble,
\begin{equation}
 R_{\rm xfit}=\frac1{m_{\rm eval}}\sum_{a=1}^{m_{\rm eval}}
\|P_{\rm xfit}u_a^{\rm(eval)}\|_2^2.
\label{eq:app-crossfit}
\end{equation}
This separation prevents the same tangent samples from both selecting and evaluating the subspace. The frozen operational profile uses 128 alignment tangents and 128 independent evaluation tangents per circuit; the generic Haar-$U(4)$ cells use 16 circuits per size and the $U(1)$ cells use 20, with a $10,000$-shot finite-shot evaluation model and 8 held-out finite-difference directions. The alignment sample is therefore a separate calibration resource: the fixed-shot comparison isolates the effect of readout orientation at fixed evaluation resources, but it is not an end-to-end claim that learning the aligned projector is free. A hardware implementation should account separately for the shots or derivative queries required to estimate $P_{\rm xfit}$.  By contrast, the same-sample Ky Fan quantity
\begin{equation}
 R_{\rm KF}=\sum_{j=1}^r\lambda_j(\widehat C)
\end{equation}
is an optimistic spectral upper benchmark rather than the operational estimator \cite{KyFan1949}.

The independent resampling unit is the circuit instance.  Tangent directions generated from the same circuit are correlated and are therefore not treated as independent bootstrap replicates.  Confidence intervals reported in the manuscript resample circuit instances and recompute the relevant aggregate.  The frozen profiles, seeds, shard-level outputs, and paper-facing summaries are archived in the reproducibility repository \cite{AitHaddou2026MeasurementAccessible}.

\section{Operational signal diagnostics}
\label{app:operational}

Let $F_{\rm full}(v)$ denote the classical Fisher scale of a direction before the unit normalization used to construct $u_v$.  For a rank-$r$ projector $P$, the raw retained directional Fisher energy is
\begin{equation}
 E_{\rm raw}(P)=\mathbb E_v\!\left[F_{\rm full}(v)\,\|Pu_v\|_2^2\right].
\label{eq:app-raw-energy}
\end{equation}
If a scalar readout direction is sampled isotropically inside the retained $r$-dimensional subspace, the mean squared directional signal is
\begin{equation}
 \mathbb E[g_{\rm dir}^2]=\frac{E_{\rm raw}(P)}{r}.
\label{eq:app-dir-signal}
\end{equation}
The finite-shot calculation in the numerical campaign combines this signal with the multinomial covariance of the fixed measurement record.  All orientation comparisons keep the circuit, quantum measurement, readout rank, and shot count fixed.  Consequently, ratios between physical, random-rank, and aligned readouts isolate the effect of the retained score-space orientation.  These quantities are controlled directional diagnostics and are not claims about the gradient variance of an arbitrary supervised loss.

\begin{figure*}[t]
\centering
\includegraphics[width=.97\textwidth]{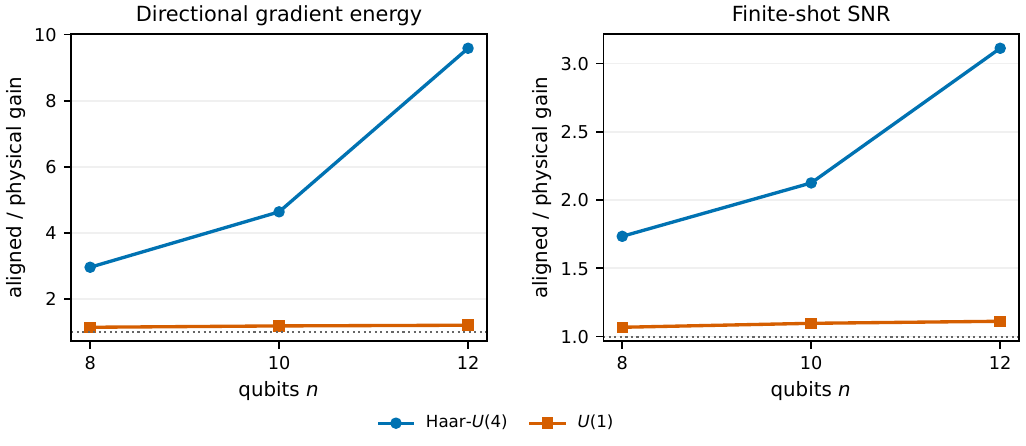}
\caption{Additional fixed-rank operational diagnostics.  Retention, directional gradient-energy, and finite-shot SNR gains are shown for the same physical versus cross-fitted aligned comparison used in the main text.  Generic Haar-$U(4)$ circuits leave substantial room for orientation improvement, whereas the structured $U(1)$ readout is already comparatively aligned.}
\label{fig:app-operational}
\end{figure*}

\section{Additional covariance and architecture diagnostics}
\label{app:architecture}

The main text emphasizes the coexistence of near-rank-typical retention with strong anisotropy.  The following supplementary views expose the corresponding effective-dimension and orientation diagnostics directly.

\begin{figure*}[t]
\centering
\begin{minipage}{.32\textwidth}\centering
\includegraphics[width=\linewidth]{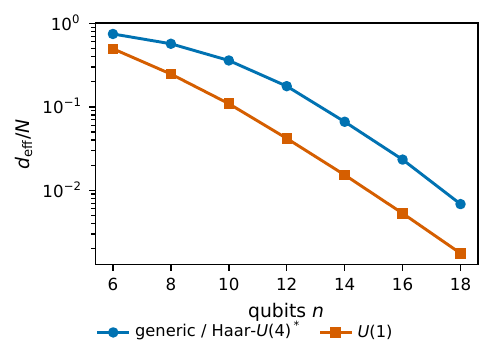}
\end{minipage}\hfill
\begin{minipage}{.32\textwidth}\centering
\includegraphics[width=\linewidth]{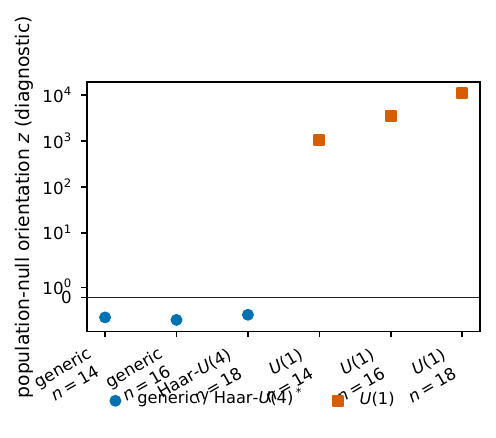}
\end{minipage}\hfill
\begin{minipage}{.32\textwidth}\centering
\includegraphics[width=\linewidth]{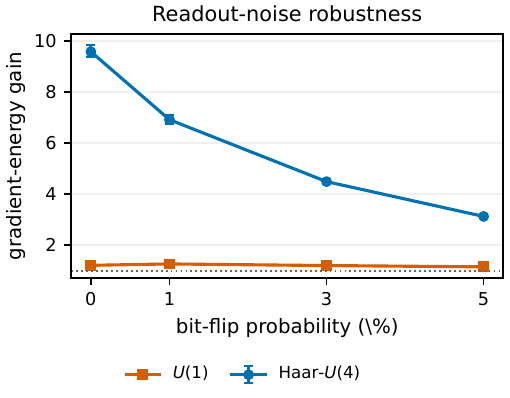}
\end{minipage}
\caption{Supporting diagnostics.  Left: effective spectral fraction $d_{\rm eff}/N$.  Center: population-null orientation diagnostic for representative generic/Haar and $U(1)$ points.  Right: aligned-to-physical directional gradient-energy gain under readout bit-flip noise.  The noise comparison re-estimates the aligned subspace after noise is applied and therefore does not establish robustness of a single fixed optimized observable.}
\label{fig:app-supporting}
\end{figure*}

\section{Finite-size diagnostics for the half-filled $U(1)$ family}
\label{app:u1_scaling}

The main text reports the cross-fitted overlap through $n=18$ and compares two simple finite-size models.  We stress that this comparison is descriptive model discrimination over the simulated window rather than an asymptotic theorem.  The fitted power exponents and $\Delta\mathrm{AICc}$ values are computed from the archived paper-facing summaries \cite{AitHaddou2026MeasurementAccessible}.

\begin{figure*}[t]
\centering
\includegraphics[width=.95\textwidth]{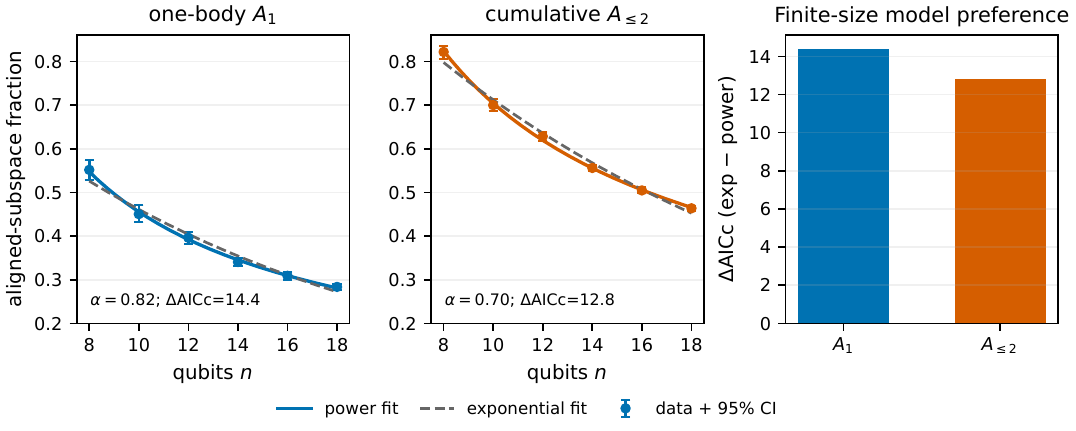}
\caption{Finite-size diagnostics for the half-filled $U(1)$ family.  The left panels show the full $n=8$--$18$ overlap data together with the fitted power and exponential models; the right panel summarizes the model-selection difference $\Delta\mathrm{AICc}=\mathrm{AICc}_{\rm exp}-\mathrm{AICc}_{\rm power}$.  Positive values favor the power model within the tested finite-size window only.}
\label{fig:app-u1-fits}
\end{figure*}

\section{Circuit conventions and reproducibility}
\label{app:circuits_repro}

The PennyLane diagrams in the main text are schematic: parameter values are intentionally suppressed, and only a few layers are drawn for legibility.  The simulations use the exact gate constructors, depth conventions, deterministic seeds, and frozen profiles archived in the repository \cite{AitHaddou2026MeasurementAccessible}.  For reference, the individual generated circuit diagrams are collected below.

\begin{figure*}[t]
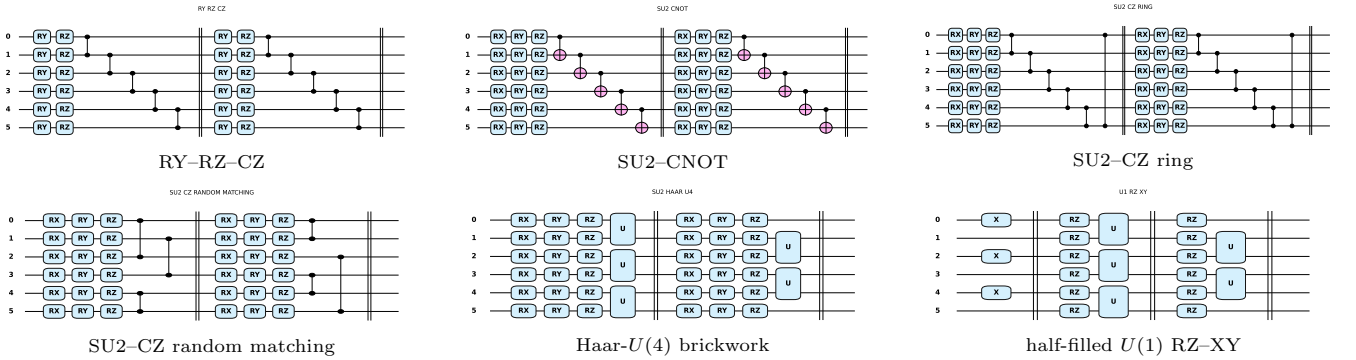

\centering
\begin{minipage}{.32\textwidth}\centering
\includegraphics[width=\linewidth]{generated_circuits/circuit_ry_rz_cz.pdf}\\[-1mm]{\scriptsize RY--RZ--CZ}
\end{minipage}\hfill
\begin{minipage}{.32\textwidth}\centering
\includegraphics[width=\linewidth]{generated_circuits/circuit_su2_cnot.pdf}\\[-1mm]{\scriptsize SU2--CNOT}
\end{minipage}\hfill
\begin{minipage}{.32\textwidth}\centering
\includegraphics[width=\linewidth]{generated_circuits/circuit_su2_cz_ring.pdf}\\[-1mm]{\scriptsize SU2--CZ ring}
\end{minipage}

\vspace{2mm}
\begin{minipage}{.32\textwidth}\centering
\includegraphics[width=\linewidth]{generated_circuits/circuit_su2_cz_random_matching.pdf}\\[-1mm]{\scriptsize SU2--CZ random matching}
\end{minipage}\hfill
\begin{minipage}{.32\textwidth}\centering
\includegraphics[width=\linewidth]{generated_circuits/circuit_su2_haar_u4.pdf}\\[-1mm]{\scriptsize Haar-$U(4)$ brickwork}
\end{minipage}\hfill
\begin{minipage}{.32\textwidth}\centering
\includegraphics[width=\linewidth]{generated_circuits/circuit_u1_rz_xy.pdf}\\[-1mm]{\scriptsize half-filled $U(1)$ RZ--XY}
\end{minipage}
\caption{Individual PennyLane circuit schematics used by the numerical campaign.  Gate parameters are hidden in the drawings, while the source constructors and complete numerical profiles are archived with the code and results.}
\label{fig:app-circuits}
\end{figure*}

\begin{acknowledgments}
The author thanks Mohamed Bennai\,\orcidlink{0000-0002-7364-5171} for helpful discussions. ChatGPT (OpenAI) was used for language editing and manuscript revision. The author reviewed the scientific content and approved the final text.
\end{acknowledgments}

\section*{Data and code availability}
The frozen protocols, source code, aggregate tables, shard-level outputs, analysis scripts, and the numerical results reported in this manuscript are available in the public project repository \cite{AitHaddou2026MeasurementAccessible}. The repository records the frozen numerical provenance separately from editorial manuscript changes.


\end{document}